\documentstyle[12pt,psfig]{article}
\def\beq{\begin{equation}}   \def\eeq{\end{equation}}
\newcommand{\bq}{\begin{eqnarray}}   
\newcommand{\eq}{\end{eqnarray}}
\newcommand{\ra}{\rightarrow}
\begin{document}
\begin{titlepage}

\begin{flushright}
NYU-TH-00/06/01\\
OUTP-00-26P\\
TPI-MINN-31/00\\
UMN-TH-1910\\
hep-th/0006213
\end{flushright}

\vspace{1cm}

\begin{center}
\baselineskip25pt

{\Large\bf Topological Effects in Our Brane World From Extra 
Dimensions}

\end{center}

\vspace{0.3cm}

\begin{center}

{\large G. Dvali \\}
\vspace{0.1cm}
{\it Department of Physics, New York University, New York, NY 10003, 
USA \\}
\vspace{0.1cm}
{\large Ian I. Kogan \\}
\vspace{0.1cm}
{\it  Theoretical Physics, Oxford University, 1 Keble Road, Oxford 
OX13NP, UK\\}
\vspace{0.1cm}
{\it and\\}
\vspace{0.1cm}
{\large  M. Shifman \\} 
\vspace{0.1cm}
{\it  Theoretical Physics Institute, University of Minnesota,
Minneapolis, MN 55455, USA}

\vspace{2cm}

{\large\bf Abstract} 

\vspace*{.25cm}

\end{center}

 The theories in which our world presents a domain wall 
(brane) embedded in large extra dimensions
predict new types of topological defects. These defects arise due to the
fact that
the  brane on which we live 
 spontaneously breaks isometries of the extra space
giving mass to some graviphotons.  In many cases the
corresponding 
vacuum manifold has nontrivial homotopies -- this
gives rise to
topologically stable defects 
in four dimensions, such as cosmic strings and monopoles that carry
gravimagnetic flux.  The core structure of these defects is somewhat
peculiar.  Due to the fact that the translation invariance in the
extra direction(s) is restored in their core, they act as 
``windows" to  the extra dimensions.
We also discuss the corresponding 
analog of the Alice strings.  Encircling such an
object one would get transported onto a parallel brane.

\end{titlepage}

 \section{Introduction.}

In the conventional  Kaluza-Klein approach \cite{1}
the Universe has a topology
$M_4\otimes K$, where $M_4$ is  our four-dimensional
Minkowski space and $K$ is some compact manifold, with the volume
typically set by a fundamental Planck length $l_{\rm P_{f}} =
1/M_{\rm P_{f}}$.  Isometries of $K$  are then seen as gauge 
symmetries of an
effective four-dimensional theory, with the role of the gauge fields
played by extra
 components of the graviton 
(the so-called graviphotons).  In the theories where our world
is a domain wall (brane) embedded in the extradimensional space
\cite{RS,DS} the situation 
drastically changes.  First, the domain wall
allows one \cite{add} to choose  a  much larger size of
the  extra dimensions,   $R \gg l_{\rm P_{f}}$.
  Second, since the branes are localized in $K$, they spontaneously break 
all
or a part of the isometries of $K$; the  corresponding graviphotons get
masses. In our
 four-dimensional world this breaking is seen as a Higgs effect,
where the role of the Goldstones  eaten up in the Higgs mechanism
is played  by the
zero modes of  the broken translational invariance \cite{add1, DS1}.  

 (In addition
in the presence of branes {\em and} gravity, 
the  space-time is not  a
direct product $M_4\otimes K$  strictly speaking, due to the 
gravitational
 field of the
brane. However, at least for the spaces with the co-dimension $N\geq 
2$, this
effect is of little importance for the present purposes, and  can be 
ignored. An  example of a solution including gravity
which  describes a
single brane on   $M_4\otimes S_1$ is given in \cite{kkop}.)

 The purpose of the present work is to study  possible macroscopic
topological 
consequences of the above picture for four-dimensional physics. 
Our starting point is the following four observations on which we would 
like to
elaborate.

$\bullet$ If topology of $K$ is nontrivial and so is that of the target 
space
${\cal T}$ (i.e. the space of  fundamental fields\footnote{These 
``fundamental fields"
must not to be confused with the low-energy four-dimensional fields;
rather the ``fundamental fields" are those of which the domain wall is built.
In fact, they need not be ``fields"
since one can consider the emergence of the brane
in a wider context of, say, string theory. If the original set-up is 
supersymmetric,
the mapping $K\to {\cal T}$ may or may not preserve a part of 
supersymmetry
\cite{DS,HLS}. In the former  case the low-energy theory
of the moduli fields on the domain wall is supersymmetric. In the latter case
this theory is nonsupersymmetric. Of particular interest is the case
when the mapping with the unit topological charge is
BPS-saturated,
while those with higher topological charges are non-BPS \cite{GDMS}.}),
there emerge topologically nontrivial mappings $K\to {\cal T}$ which 
are
characterized by various moduli. The moduli become dynamical fields of 
the
low-energy four-dimensional theory.

$\bullet$ The number of the moduli is equal to or larger than the 
number of
the  symmetries of the theory
 broken by the mapping under consideration. If the
topological number of the mapping $K\to {\cal T}$ is larger than one, 
this gives rise to
``parallel" branes and the ``horizontal" proliferation of the low-energy
four-dimensional fields (multiple generations). Even for the unit
topological number of the mapping $K\to {\cal T}$, the number of the
 moduli may be
significantly larger than the number of the broken isometries.
This phenomenon corresponds to dynamical  symmetries
and their Goldstones.

$\bullet$  Topology of the moduli space is typically nontrivial too.
In other words, the space of the effective low-energy four-dimensional
fields is topologically nontrivial. This generates physically observable 
topological
defects in our world. 

$\bullet$ Inclusion of gravity and its interplay with nontrivial  topology 
 leads to peculiar effects.

\vspace{0.2cm}

In this work we will focus on those moduli that correspond to the 
spontaneously broken isometries of $K$.
 Let $G$ be an
isometry group of $K$. A  given domain wall breaks
 this group down to a subgroup
$H$. Then $G/H$ defines a vacuum manifold; if homotopies of this
manifold are nontrivial, the theory  admits topologically nontrivial 
stable configurations.
This structure has a simple geometric meaning, as seen
directly from  the high-dimensional
Universe. Assume, for definiteness, that the brane at hand  is a
three-brane. Then it is a point on $K$. The space of all possible
locations of the brane is $K$ itself.  Now imagine that $\pi_{n}(K) \neq 
0$.
In other words,  $K$
contains $n$-dimensional closed surfaces that can not be contracted to a
point in $K$.  

For $n= 1 $ and 2,
such surfaces can be mapped on the spatial boundary of our $M_4$. 
Such
configurations will be topologically nontrivial; they can not be deformed
 to a trivial 
vacuum continuously.  They
correspond to  a constant change of the position of the brane on $K$ as 
we
travel around a closed surface in 3+1 dimensions. For instance, at $n
= 1$ the configuration we will deal with is a cosmic string. 
Winding  around such a string,
the four-dimensional  observer will  make
 a full circle on $K$ along the extra dimension.

An intriguing question is what is the core structure of such 
defects? Usually the broken symmetries get  restored (at least, partially)
 in the core
of the topological  defects. The symmetries in question  are
the  translations in the
extra space. Their restoration would mean that there is a ``hole" in
the domain wall  -- in a sense, the topological defects open a door
 in the extra space and
 can link together ``parallel"  brane worlds.

The organization of the paper is as follows.
In Sec. 2 we discuss strings, Sec. 3 is devoted to monopoles, 
Sec. 4 deals with the Alice strings,
in Sec. 5 we briefly discuss proliferation of moduli
(some of them may be related to dynamical rather than geometrical
symmetries). Finally, the  graviphoton mass
 is considered in  Appendix.

\section{ The Kaluza-Klein Cosmic String}

In this section we consider the simplest defect of the type discussed
above, the
Kaluza-Klein cosmic string. Although this structure has nothing to
do with the classical  Kaluza-Klein set-up \cite{1} -- it  exists only  in
the theories with the branes -- we keep  the name `` Kaluza-Klein,"
since this string
carries a magnetic  flux of the graviphoton 
field. The  magnetic  flux  connects the Kaluza-Klein monopoles.

The simplest possibility  with nontrivial $\pi_1$
is to assume that $K = S_1$. 
This has the isometry group $G = U(1)$, broken by the brane down to
identity. The brane  position  on $S_1$ is parametrized by one
scalar modulus $y$. This position can slowly vary with $x_{\mu}$,
the four-dimensional space-time point. Thus, the four-dimensional 
observer will perceive $y (x_{\mu})$ as
a low-energy scalar  field on $M_4$. The target space of $y$ is 
obviously a circle $S_1$. The corresponding  fundamental group
$\pi_1(S_1) = Z$. 

We will be interested in the configurations in which  the
brane sweeps  a full circle around $K= S_1$  as we travel along a
four-dimensional closed path at infinity. Let $0\leq  y < L$ (the points 
$0$ and $L$
are identified), $t$ is time of $M_4$, 
while $r, \theta$  and $z$  are the spatial 
 coordinates on $M_4$ ($r$ and $\theta$
are the polar coordinates). The Kaluza-Klein cosmic string oriented along 
the $z$
axis then corresponds to the following asymptotic configuration
\begin{equation}
\theta =  {2\pi n y \over L}
\label{string}
\end{equation}
where $n$ is an integer, the winding number. 

In the absence of gravity, such
configuration  has a logarithmically  divergent energy at large
$r$. This divergence comes from the long-range gradient energy of the
Goldstone field living on the brane.  At distances $\gg L$ the only
relevant degree of freedom describing the brane dynamics is the
Goldstone mode
$\chi \equiv  \sqrt{T}\, y(x_{\mu})$ of the broken translational 
invariance.
Here $T$ is the brane tension (energy per unit three-surface). 
The low-energy  Lagrangian obviously  has the form 
(modulo higher derivatives)
\begin{equation}
{\cal L}_{\rm eff} =  T\, (\partial_{\mu}y(x) )(\partial^{\mu}y(x) )\,.
\end{equation}
 
It is clear that the configuration (\ref{string}) corresponds to the
winding of the Goldstone field and results in the logarithmically 
divergent
energy per unit  length of the string,
\begin{equation}
E \sim L^2\, T\, n\, {\log} \rho\,,
\end{equation}
where $ \rho$ is the  maximal  distance in the direction
perpendicular to the $z$ axis along which the string is aligned,
and $L$ is the size of the fifth dimension. 
Such 
logarithmically divergent energy is typical for  global cosmic strings.
In the
 case at hand it simply indicates that in the absence of gravity
the Kaluza-Klein cosmic strings would be global \cite{global}. 

However, in actuality this divergence is
compensated by the graviphoton
field, which takes a pure gauge form at infinity,
\begin{equation}
A_{\theta} =  {n \over g}\,,
\label{aaa}
\end{equation}
where $g = 1/(M_{\rm P_{f}} L)$ is an effective gauge coupling.  
The topological defect we deal with here is
of the type of  a local
Abrikosov-Nielsen-Olesen string \cite{Abrikosov}. 

We pause here to make an important remark.
In the topologically trivial sector the graviphoton gets mass
through the Higgs mechanism -- the spontaneous breaking of  U(1).
Say, if the original space is five-dimensional,  the 
graviphoton mass squared  is proportional to the brane tension
and inversely proportional to the four-dimensional Planck mass 
$M_{p}^2
 = M_{\rm P_{f}}^3\,  L$, 
(details  are given  in Appendix)
\begin{equation}
  M^2_V = \frac{T}{M_{p}^2}\,. 
\end{equation}
Analogous formulae can be obtained for general p-branes in the
D-dimensional space-time. 
Although the above conclusion has been reached in
an effective field theory, it is 
 quite general and must be applicable to D-branes 
in string theory too --  in the
 presence of D-branes graviphotons will become    
massive. In the  weak coupling limit
 the D-brane  tension is very large, it grows   
as $T \sim M_{\rm str}^4/g_{\rm str}\,$ and, at the same time,
  the Planck mass  scales as $M_{p}^2 \sim M_{\rm
str}^2/g_{\rm str}^2\,$. Assembling these factors we obtain
that the 
 graviphoton mass
scales as
\begin{equation}
  M^2_V  \sim g_{\rm str} M_{\rm str}^2\,.
\label{nonana}
\end{equation}
 The fact that this mass is proportional to $ g_{\rm str}$, rather than  to
$ g_{\rm str}^2$, is a pure D-brane effect; it is
 due to the fact that the 
Born-Infeld action describing the D-brane dynamics is proportional to
 $1/ g_{\rm str}$.

\vspace{0.2cm}

Returning to the field configuration
(\ref{string}), (\ref{aaa}), we observe that
 the
magnetic flux is  trapped in the core of the  Kaluza-Klein string. 
 Therefore,  such strings
must be able to end on the Kaluza-Klein monopoles. This leads 
us to the conclusion
that in the  brane scenarios the
 Kaluza-Klein monopoles are not  stable; rather, they 
 get connected by the  Kaluza-Klein cosmic
strings and annihilate. 

Infinite isolated cosmic strings can also exist. At large
distances  the behavior of 
 the Kaluza-Klein cosmic strings is similar to the conventional gauge 
strings
in U(1) theories. The
core structure of these objects is
somewhat peculiar, however. 

Normally the order parameter  responsible for
the symmetry breaking must vanish in the core where the U(1) 
symmetry
 gets
restored. However, in the  case at hand the order parameter is the 
position of the
brane on $K=S_1$. Thus,
we expect  a ``hole" in the brane at the location of the
defect. This hole is a domain on $M_4$  where the translational 
invariance
on $K$  is restored. In the case at hand this is the $z$ axis.
This domain acts as a ``window" in the extra dimensions. 

In the
 case when the brane is a topological soliton of the type considered in 
Ref.~\cite{HLS}
the nature of the hole is easy to visualize. Let us discuss, for instance, 
the
theory of  one fundamental field $\Phi$,
with a non-simply
 connected target space.  One can consider, for instance,  a
real  scalar field
$\Phi$ defined modulo $2\pi$ on $M_4\otimes S_1$.
Assume that the potential is
\begin{equation}
V = \frac{1}{2}\left( \frac{C}{1+\beta\cos\Phi}\right)^2\,,\qquad C=
\frac{2\pi}{L}\,,
\label{potential}
\end{equation}
and the constant $\beta$ is positive and slightly less than 1.
 This theory admits a stable soliton solution defined by the equation
\begin{equation}
\frac{2\pi}{L} \left(y-y_0 +\frac{L}{2}\right) = \Phi + \beta\sin\Phi\, .
\label{brsol}
\end{equation}
Here $y_0$ is the soliton center. The solution 
$\Phi (y) $
 interpolates between $\Phi= 0 $ and $\Phi=2 \pi $
as $y-y_0$ varies from $-L/2$ to $+L/2$. 
By choosing the parameter $\beta$ sufficiently close to 1, one 
regulates the width of the brane in the $y$ direction making it as 
small as it is
desired.
 Since $\Phi$ is
periodic such solitons are perfectly compatible with  compactness of $K$.

 Now let us  make $y_0$ a slowly varying function of $x_\mu$.
The  cosmic string lying along the $z$ axis
is formed if $y_0$  changes from 0 to $L$ as
we wind around the
$z$ axis in the perpendicular plane. 
If we are sufficiently far from the $z$ axis
the brane soliton has the form following from Eq. (\ref{brsol}).
However,  at $r\to 0$ the  location of the brane on the  circle $K$ is 
ill-defined. This means that  at $r\to 0$ the fundamental field 
configuration
continuously evolves from (\ref{brsol}) to
\begin{equation}
\frac{2\pi}{L} \left(y-y_0 +\frac{L}{2}\right) = \Phi \, .
\label{brsolprim}
\end{equation}
The latter corresponds to the brane completely smeared over $K$,
so that $y_0$  looses its meaning of the brane center.
The width of the brane in the $K$ direction becomes $L$.
The field configuration (\ref{brsolprim}) has an excess of potential 
energy
-- this is standard for the core of the  string.
An observer approaching the cosmic string will leave our brane
and fill $K$ entirely.  

The whole construction can be viewed as a
circular ``staircase'' -- split  our three-dimensional space  in a sequence
of two-dimensional planes attached to the given string; passing from one 
plane
to another we simultaneously shift   in the  fifth direction, as  shown in 
Fig. 1.
In fact, the picture is somewhat more contrived, since the slices
$y=0$ and $y=2\pi R\equiv L$ must be identified,
but this is impossible to show in the figure.

\begin{figure}
\centerline{\psfig{file=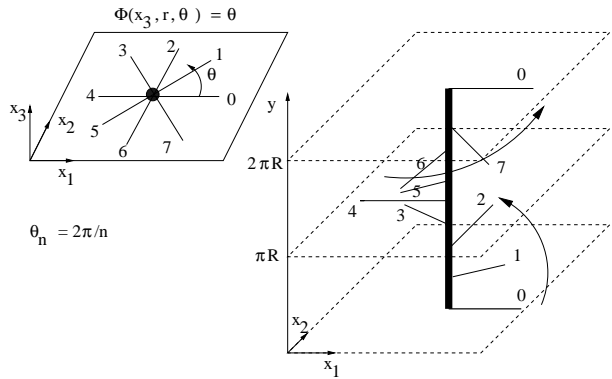,width=8cm}}
\caption { $K=$U(1)  Kaluza-Klein string as a  ``staircase.''  
Our three-dimensional
space is  represented as an ensemble of planes attached to the string
which lies along the $z$ axis
(see a slice on the left).
As we move from one plane to another we
simultaneously shift in the fifth direction. A  full winding corresponds
to the shift from $y=0$ to $y=2\pi R\equiv L$.
}
\label{Fig.1}
\end{figure}

If originally there were several branes on $K$ (this would correspond to 
higher
windings of $\Phi (y)$)
at $r\to 0$ the proper field configuration must
evolve to 
\begin{equation}
\frac{2\pi \nu}{L} \left(y-y_0 +\frac{L}{2}\right) = \Phi \, ,
\label{brsolpr}
\end{equation}
where $\nu$ is the winding number. Near the position of the cosmic 
string
all $\nu$ branes fuse together. The construction with several ``parallel"
branes may be promising from a phenomenological standpoint
\cite{GDMS}.

\section{The Monopoles}

Let us  now pass to the discussion of 
nontrivial $\pi_2 (K)$.
 We will show that the brane can create certain topologically stable
configurations which look as monopoles in four dimensions.
Again, we will focus on the simplest manifold with
nontrivial $\pi_2 $, 
the  two-sphere $K= S_2$. The brane breaks the isometries of $K$.
The symmetry breaking pattern on $K$ is
SU(2) $\to$ U(1). Correspondingly,
the position of the brane 
on $S_2$ can be characterized by two angular coordinates
$0 < \theta_{K} < \pi$ and $0 < \phi_K < 2\pi$. 
In the low-energy four-dimensional theory
$\theta_{K} (x_\mu )$ and $\phi_{K} (x_\mu )$
become the Goldstone fields.
To parametrize $M_4$, instead of
$t,r,\theta , z$ of Sec. 2 we now 
introduce the spherical coordinates $r, \theta, \phi$. Since 
the brane can
fluctuate and move on $S_2$, the angles $\theta_{K}$ and $\phi_{K}$ are
slowly varying functions  of $r, \theta, \phi$ and 
our Minkowski time $t$.
The topologically stable
monopole configuration is given by the following mapping:
\begin{equation}
\theta_{K} = \theta,~~~ {\rm and} ~~~\phi_K = \phi\,,
\end{equation}
at  $r $ sufficiently far from the monopole core,
plus the appropriate configuration of the graviphoton fields. 
The general structure is the same as that of the 't Hooft-Polyakov
monopoles \cite{THP} in the Georgi-Glasow model.
This configuration
carries a topological charge since it corresponds to the mapping $(S_2)_K
\rightarrow (S_{2})_{M_4}$, where the second sphere is the
 spatial  boundary of the spatial part of 
$M_4$.

At $r$ approaching zero
the symmetry must be restored, and the ``former" brane must delocalize 
on $K$
much in the same way as in the cosmic string example (Sec.~2).
Thus, the core of the monopole presents an exit into $K=S_2$.
If there are several ``parallel" branes, the monopole core 
will connect all of them. In a sense, this is even a more interesting object 
than
the string of Sec. 2 since it is fully localized on $M_4$.
The mass of such  monopole is expected to be of order
\begin{equation}
M_{\rm monopole} \sim (\sqrt{T}R^2) \, M_{\rm P_{f}}\,,
\end{equation}
where $R$ is the radius of $S_2$. 
The monopole mass can be much lighter than the fundamental Planck 
scale,
since in our scenario the product $\sqrt{T}R^2$ depends on 
dynamical details of the
underlying theory, and can well be small. Finding
this monopole would be an exciting endeavor since one could channel
signals to/from other branes  through its
core.

\section{The Alice Strings from the  Branes}

In this section we will deal with $K = S_2$,
but nontrivial $\pi_1$, rather than $\pi_2$. 

\vspace{0.2cm}

 In the example of Sec. 3, the brane was represented by a point
on $K = S_2$. What happens if the  brane
 is represented by two or more {\it rigidly}
connected points? 
For instance, assume  the brane to be  described by two
identical points in  the opposite poles of the two-sphere.
After one of the isometries of $S_2$ is broken, 
the surviving symmetry then is $U(1)\times Z_2$, since one
can interchange the  poles without affecting the  
expectation value of
the brane position moduli 
on $K$. 
Strictly speaking, the surviving symmetry is not the
 direct product, however, since
$Z_2$ flips the sign of the $U(1)$ generator. This sign-flipping
results in the existence
of  Alice strings \cite{alice} in the four-dimensional space $M_4$. 

To see that  this is indeed the case, let the surviving U(1)
generator be $\tau_3$ (the third Pauli
matrix). Then $Z_2$ can be represented as
$$
U_{Z_2}= \exp\{-(i/2) \, \pi\,  \tau_2\}\,,
$$
so that $U_{Z_2}^{-1} \,\tau_3 U_{Z_2} = - \tau_3$.
 The vacuum manifold contains
unshrinkable paths which start at identity and end at $U_{Z_2}$. 
This can be parametrized by a group transformation
\begin{equation}
U_{\vartheta}= \exp\{-(i/2) \, \vartheta \,  \tau_2\}\,,\quad
\vartheta \in [0,\pi]\,.
\end{equation}
The Alice string configuration is  obtained by identifying $\vartheta$ 
with
the  angle $\theta /2$ in $M_4$ introduced in Sec. 2. In other words, as 
one winds
once around the $z$ axis in $M_4$ (far away from the axis)
the position moduli on $K$ drift from the north to the south pole
(which are identified).
This configuration is obviously topologically stable. 

 In this way the
opposite points on
$K$ interchange places when one travels once around the string (far 
away 
from the string 
axis). If such observer completes his/her journey
and   comes back to the point of departure, he/she will find
U(1) charges to be conjugated. Indeed, assume the travel was
adiabatic. The wave function of any state will
track observer's position
so that the wave function will
acquire  the  gauge factor
\begin{equation}
\psi(\theta ) = \left( U_{\vartheta}\right)^q \psi(0)\,,
\end{equation}
where $q$ is the U(1) charge of the state under consideration measured
in the units of the fundamental charge.
One can measure the U(1) charge at the beginning
and the end of the journey by acting by $\tau_3$ on  states
$|\psi(0)\rangle$ and $|\psi(2\pi)\rangle$, respectively,
 assuming the travel to be
adiabatic.
 However, 
$$
|\psi(\theta= 2\pi )\rangle
= \left(U_{Z_2}\right)^q\, |\psi(0)\rangle\,.  
$$
For all odd $q$ 
$$
\tau_3 |\psi(\theta= 2\pi )\rangle = (-1)\, |\psi(\theta= 2\pi )\rangle\,,
$$
while for even $q$
$$
\tau_3 |\psi(\theta= 2\pi )\rangle = (+1)\,| \psi(\theta=
0 )\rangle\, .
$$
 For instance, if we
transport a fermion in the fundamental representation of 
the original
SU(2) around the
string, its U(1) charge will flip the sign. This is a typical behavior
for the  Alice
strings.

In the present case the emerging  Alice
string has a clear-cut geometric interpretation. The conserved U(1)
is nothing but the rotation of $S_2$ around the axis connecting the north 
and
south poles. 
Traveling around the  string 
(i.e. winding around the $z$ axis in $M_4$)
 interchanges the north/south poles and
thus, the angular momentum of any state. One can visualize
this as a loop made of a M\"obius strip. 

\begin{figure}
\centerline{\psfig{file=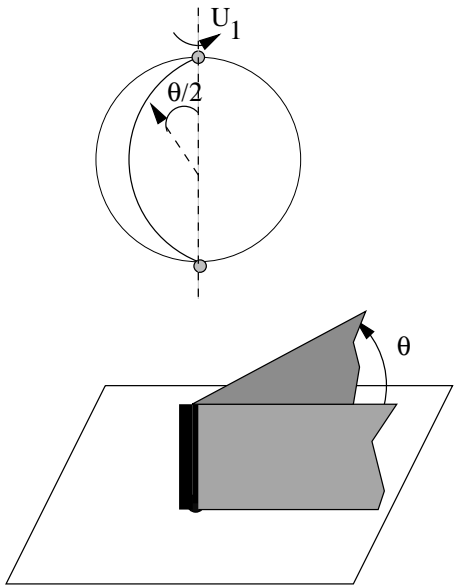,width=6cm}}
\caption {$S^2$ with two branes at antipodal points. The Alice string is
given by the  large semicircle connecting these two points. The
(3+1)-dimensional space is distributed all over the  semicircle in
such a way that a  plane attached to the Alice  string at angle $\theta$ 
is connected  to the point on the semicircle parametrized by
the  angle $\vartheta = \theta/2$.
Making a full $2\pi$ rotation of the  plane we  one  moves from
the north to the south pole on $S^2$ and  by this changes the sign of the
U(1) generator.}
\label{Fig.2}
\end{figure}

In a similar manner one can 
 get   strings which  interchange two gauge groups (or possibly 
 more than two). Let us consider, for example, the case when
the original gauge symmetry  $G$ (related to the isometries of $K$)
is broken down
to a subgroup
$H \times H$.
 We want to have a $Z_2$ subgroup of $G$ which does not commute with 
 $H \times H$ in order to generate  a $Z_2$ string such that after making 
a full
turn we exchange two $H$'s. This will have a very spectacular effect 
--
if such a string  passes between a source and a detector measuring
the flux of charged  particles
(charged  with respect to  one of the gauge  groups only), the detector 
 will see
only a half of the beam. The reason that  half of the particles will be
transformed into particles charged with respect to a second group and
our detector will not see them. What half will undergo the transition
depends on the history of how the  detector and  the source were
 prepared.

How to make such a string? Take the manifold $G/(H\times H)$. By 
definition there is a point $A$ on this manifold which is stable under
$H \times H$ -- so if we put a brane at this point  the  $H \times H$
subgroup of $G$ remains unbroken. 
Consider
$G$ as an ensemble of $2n \times 2n$ matrices and $H$ as $n \times n$. 
Take now the group element
\begin{eqnarray}
 T = \left( \begin{array}{cc}
0 & I\\
 I & 0\\ \end{array}\right)
\end{eqnarray}
where $I$ is the $n \times n$ identity matrix. 
One can see that $T ^2 = 1$ while $T$  interchanges 
two $H$ subgroups. The  antipodal point $B$ 
is constructed  by acting by $e^{i\pi T}$ on $A$. We then  
  consider the coset   $G/(H\times H)$ with the  branes placed at both 
points
$A$ and $B$. Besides the unbroken symmetry $H \times H$ the group 
$Z_2$ is also
unbroken  -- it interchanges $A$ and $B$ (with one brane this
symmetry would be  broken).
Moreover, this $Z_2$   does not commute with  $H \times H$.
The Alice string is obtained  by splitting our space in planes
and attaching them along the  line $e^{i\theta T/2}A$ for $\theta$
between $0$ and $ 2\pi$.   In principle, one may try to consider $Z_N$
strings by invoking  $H \times .. \times H$ symmetry with $N$ copies
of $H$ and the coset $G/(H \times .. \times H)$.

(Let us parenthetically note
that  there may occur  obstructions to construction of the  Alice strings,
see e.g. \cite{Mcinnes}. 
Classification of  all possible Alice strings within the
brane scenarios is an interesting question.)

\section{Proliferation of Moduli}

The effective theories considered so far were
those of the geometric moduli related to the isometries of $K$.
The number of moduli can be much larger, however,
since in the given fundamental theory
responsible for the brane formation symmetries may be 
dynamical. 
The nontrivial topology of the moduli space can be much more
contrived than that of $K$.
Here we will discuss a simple example of this phenomenon.

Assume that $K=S_2$ and the underlying theory
is the nonlinear O(3) sigma-model, with the fundamental field
 $\vec{n}$ where $  \vec{n} ^2 = 1$. 
The combined action, including gravity,
is 
\bq
S =  M_{\rm P_{f}}^4
\int d^4 x d^2y \sqrt{G} R^{(6)} + \int d^4 x d^2y \sqrt{G}\left[
\frac{1}{2} G^{MN}\partial_M\vec{n}\partial_N\vec{n} 
\right]+ ...
\label{6action}
\eq
where $M_{\rm P_{f}}$ and $R^{(6)}$   are the six-dimensional 
fundamental
Planck 
mass and the scalar curvature, respectively, the dots stand for the
possible fermion terms if the theory is supersymmetric, and finally
$M,N = 0,1,2,3,4,5$. 

Now we  consider classical solutions of the equations of motions 
with nontrivial topology, depending on $x_4$ and $x_5$
where $x_{4,5}$ parametrize $S_2$. These are nothing but
the Polyakov-Belavin instantons \cite{Polyakov:1975yp}.

 Using the
standard  stereographic projection
\bq
W = \frac{n_1 + i n_2}{1+ n_3}
\eq
one can introduce a complex field $W$ depending
on a complex variable $z$ ($S_2$ is the K\"ahler manifold
allowing for the complex structure)
\bq
W(z) = \prod_{i=1}^{K} \frac{z-b_i}{z-a_i}
\eq
where $z$ is the complex coordinate on $CP^{1}$, and
$K$ is the winding  number. Moreover, $b_i$  and
$a_i$ are complex numbers, $2K$ altogether,
representing the  moduli parameters  of
the soliton. 
 
 Now we can make them coordinate-dependent,
 \bq
W(z,x^{\mu}) = \prod_{i=1}^{K} \frac{z-b_i(x)}{z-a_i(x)}\,.
\eq
The action can be written as (note that
$\partial_z \bar{W} = 0$)
\bq
\int d^4 x \frac{dz d\bar{z}}{(1+ |z|^2)^2} \sqrt{g} \left[
\frac{|\partial_z W|^2 }{(1+|W|^2)^2} +
G^{\mu\nu}\frac{\partial_\mu W \partial_\nu \bar{W}}{(1+|W|^2)^2} +
G^{\mu z}\frac{\partial_z W \partial_\mu \bar{W}}{(1+|W|^2)^2} +\mbox{ c.c.}
\right].
\eq
We see that
\bq
\partial_z W = W \sum_{i=1}^{K}\left[\frac{1}{z-b_i}
-\frac{1}{z-a_i}\right] \,, \nonumber \\[0.2cm]
\partial_\mu W = W \sum_{i=1}^{K}
\left[\frac{ \partial_\mu a_i}{z-a_i} -  \frac{\partial_\mu b_i}{z-b_i} \right].
\eq
Substituting this in the action we  get an effective Lagrangian
 describing dynamics
 of the collective coordinates (moduli) $a_i(x)$ and $b_i(x)$,
 \bq
 S[a,b] &=& \int d^4x \sqrt{g} G^{\mu\nu}
\left[
F^{ij}(a,b) \partial_\mu a_i \partial_\nu \bar{a}_j 
\right. \nonumber
\\[0.2cm]
&+&\left.
B^{ij}(a, b)( \partial_\mu a_i\partial_\nu  \bar{b}_j
+ \partial_\mu  b_i\partial_\nu \bar{a}_j ) +
H^{ij}(a, b)\partial_\mu  b_i\partial_\nu  \bar{b}_j
\right]\,,
\label{instaction}
 \eq
 where $F^{ij}, B^{ij}, H^{ij}$ represent a metric on the  $2K$-dimensional
moduli space. We do not  know the explicit form of the
metric  on the  multi-instanton moduli 
space (i.e. at $K > 1 $). It is known, however,  that this moduli space is a 
 K\"ahler manifold.
 
In case of $S_2$ we have three graviphotons 
-- the gauge symmetry is
SU(2) -- but we have an 
 arbitrary number of the Goldstone modes
provided $K$ can be chosen
at will. If $K=1$,
the  Goldstone 
 action for the moduli $a$  and $b$ can be easily found. We can
introduce two new collective coordinates: $ X = a-b$ (the size 
of the instanton and U(1) orientation) and $Y = (a+b)/2$ (the 
position of the center). 

We will  work in the approximation when $a$ and $b$ are small in which
case  we explore only a small patch on $S_2$ which is an open region of 
the  plane. The action of  the SU(2)  generators in this limit is nothing
but  a  U(1) rotation (which will be our unbroken symmetry) and two 
translations (the non-Abelian nature of the  SU(2) is not seen
in this approximation).

The action 
(\ref{instaction}) in this case can be written as a sum of two
independent actions, for the $X$ field\footnote{
 Here the  logarithmic dependence on the size of instanton in the metric is 
related to the well-known fact that the measure on the moduli space has 
the  form $dX d \bar{X}/|X|^2$ as dictated by the  conformal
invariance of the classical action.} 
\bq
S_{X} \sim \int d^4 x \ln |X|^2 |\partial_{\mu} X|^2\,,
\eq
and for the $Y$ field
\bq
S_{Y} \sim \int d^4 x  |\partial_{\mu} Y|^2\,.
\eq 
After the inclusion of the  graviphotons
 the action  for the $Y$ fields becomes\footnote{This is   not a
manifestly SU(2) invariant expression because we neglected higher
order terms in $Y$. If one does the calculation taking into account
that the soliton  is on $S_2$, rather than on $R_2$, the full SU(2)
invariant action can be recovered.}
 $$(\partial_\mu Y + A_\mu^{+})(\partial^\mu \bar{Y}
+ A^{\mu -})\,.$$
We arrive at the
  Georgi-Glashow model SU(2) $\ra$ U(1).  Let us note that the modulus
$X$ does not mix with the graviphoton, only $Y$ does. This is because it is
$Y$ that is transformed by two broken symmetries of SU(2).

 For $K > 1$ we  have more massless particles.
 In the supersymmetric case we have  not
only bosonic zero modes, but fermionic  as well. The latter are
chiral, so  we can create several  generations of the chiral matter.  This
example shows that having a topological defect within a brane world
scenario may not only open a window to extra dimensions but can also
create a large number of light fields. To this end, the 
topological defect at hand must have a  large enough topological
charge, i.e. a large dimension of its moduli space.

\section{Conclusions}

In the theories where our world is trapped on the brane(s)
embedded in large compact extra dimensions
the Kaluza-Klein scenarios get drastically modified.
Since the brane break isometries of the
extra space $K$, the graviphotons
become massive via an analog of the Higgs mechanism.
In addition, a nontrivial topology of the compact space $K$
gets entangled with the topology of our world, giving rise
to strings (in case of nontrivial $\pi_1$) and monopoles
(nontrivial $\pi_2$) of a special geometric nature
(we call them Kaluza-Klein defects,
although they can appear {\em only}
in the presence of branes). In the core of the Kaluza-Klein defects
the full symmetry of $K$ gets restored, so that their cores
represent a natural channel of exit into $K$. If there are several 
``parallel"
branes, they may get connected through the core of the monopole or the
axis of the strings. We have considered two
types of strings -- the Abrikosov-Nielsen-Olesen string
and the Alice string.

A large number of additional moduli may (and usually do)
 naturally emerge,
which have a dynamical rather than geometric origin. 
This results in the proliferation of matter trapped on the branes.
The number of moduli is proportional to the topological number of the
mapping which need not necessarily be unity. 
The topology of the
moduli space is typically more contrived than that of $K$.
Implications of this observation for the
topological defects  observable in our world
are yet to be studied. 

The original motivation \cite{RS}
for introducing brane worlds
was the desire to localize matter in a `` transverse"
space of a small volume embedded in a noncompact space of 
the infinite volume. The latter was then replaced by a compact
space of a large size \cite{add,add1}, where the branes were supposed
to accomplish the same mission of localization.
At present the importance of  this aspect of the brane scenarios -- 
localization -- fades away, as, on the one hand, people start to realize 
that other aspects of the brane scenarios may be potentially 
instrumental, and on the other hand,  extra spaces of exceedingly 
smaller 
sizes are emerging in various theories (e.g.  \cite{Chacko:1999eb}).
Other goals which might be  achieved in the brane world
scenarios are taking over; they are: (i) supersymmetry breaking and 
separation of chiralities (i.e. making our matter chiral starting from a 
nonchiral set) \cite{DS};
(ii) hierarchies in the SUSY breaking parameters and mass parameters 
\cite{GDMS}; (iii) generation of a mass term for the graviphotons
\cite{add1,DS1}; (iv) proliferation of matter -- i.e.  the ``parallel" 
matter generations. The topological defects of the special type, discussed 
in this paper, is one more new aspect. 

\vspace{0.3cm} 
{\bf Acknowledgments}

 \vspace{0.1cm} 
We are grateful to A. Tseytlin for a useful  discussion
and to Albert
Schwarz for
useful communications. 
The work of G.D. was supported in part by David and Lucille  
Packard Foundation Fellowship for Science and Engineering.
The work of M.S.
was supported by  DOE  grant DE-FG02-94ER408. 
The work of I.K. was supported in part by PPARC rolling grant
PPA/G/O/1998/00567 and  the EC TMR grant FMRX-CT-96-0090

\section*{Appendix: Branes Make Graviphotons Massive}

\renewcommand{\theequation}{A.\arabic{equation}}
\setcounter{equation}{0}

The brane-Higgs effect has been already discussed in \cite{add1,DS1}.
Here we present it for completeness and also  give a detailed
derivation of the graviphoton mass. 
 Let us start from the simplest situation: consider a
 domain wall
 in the five-dimensional space $M^4 \times S_1$ 
with four noncompact
 coordinates $x_0, x_1, x_2, x_3$ and one compact  (fifth) 
coordinate $x_4 = y$.
Imagine that the
domain wall is  made of a scalar field $\Phi$; the combined
action including gravity  is 
\bq
S =  M_{\rm P_{f}}^3\int d^4 x dy \sqrt{G} R^{(5)} + \int d^4 x dy 
\sqrt{G}\left[
\frac{1}{2} G^{MN}\partial_M\Phi\partial_N\Phi + V(\Phi)
\right]
\label{5action}
\eq
where  $ M_{\rm P_{f}}$ and  $R^{(5)}$ 
 are the five-dimensional Planck mass and the scalar curvature, 
respectively, and 
$M,N = 0,1,2,3,4$. The  Greek letters $\mu, \nu =  
0,1,2,3$ are reserved for the  four-dimensional indices. 
The signature is $(-,+,+,+,+)$.
The explicit form of the potential $V(\Phi)$ is not important here --
the only thing we have to know is that the target space has 
a nontrivial
topology and 
there are topologically stable classical solutions.
  For example one can consider the potential (\ref{potential}).
Let us forget about gravity for a moment.
The domain wall $\Phi_0(y)$ in the  fifth direction 
is given by the solution of the classical equations
\bq
 \frac{d^2 \Phi_0}{d y^2} -  V'(\Phi_0)=0\,,
\label{classicalequation}
\eq
with the first integral
\bq
\frac{1}{2}\left(\frac{d \Phi_0}{d y}\right)^2 - V(\Phi_0)=0\,,
\eq
which gives us the
 tension (energy density in $M^4$) of the domain wall
\bq
T = \int dy \left[\frac{1}{2}
\left(\frac{d \Phi_0}{d y}\right)^2 + V(\Phi_0)\right] =
\int dy  \left(\frac{d \Phi_0}{d y}\right)^2 \, . 
\label{tension}
\eq
The solution of the  classical equation (\ref{classicalequation}) 
depends on one parameter --
 the position of the domain wall along the fifth 
direction  $\Phi_0 = \Phi_0(y - \phi R)$, where $\phi \in [0, 2\pi)$
 and $R$ is the  radius of $S_1$. 
If we  consider now the spectrum of small fluctuations of
the  scalar 
field, the parameter $\phi$ will become the
 collective coordinate in the  expansion around 
the domain wall background,
\bq
\Phi(x, y)= \Phi_0(y - R\phi(x)) + \sum_{n \neq 0} \phi_{n}(x)v_n(y)\,.
\label{expansion}
\eq
Here only non-zero modes $v_n(y)$ orthogonal  
to the zero mode $v_0(y) = (d\Phi_0/dy)$ are included
in the sum. It is easy to
see   that $\phi(x)$ is a Goldstone  field -- independently of the
form of the potential    this field will be massless at
the  quantum level. The
reason is simple -- the Goldstone     theorem guarantees 
that there  is a
massless field if a global  continuous symmetry   is broken. 
In our case we deal with a
``translational"  global U(1) symmetry
\bq
\phi \ra \phi + \epsilon\,,
\label{globalu1}
\eq
where the angle $\phi$ is related to
the  fifth coordinate $y$ as $y = R\phi$.
This symmetry
  is broken by the solution  $\Phi_0(y - R\phi(x))$. 
Correspondingly,  the zero mode
$\phi(x)$ represents   the Goldstone boson with  the
 action
\begin{eqnarray}
S[\phi] &=& \frac{1}{2}\int d^4 x dy
\partial_\mu\Phi_0\partial^\mu\Phi_0 =
\frac{R^2}{2}\int d^4 x  dy \left(\frac{d\Phi_0}{dy}\right)^2 
\partial_\mu\phi\partial^\mu\phi 
\nonumber\\[0.2cm]
&=& \frac{f_\phi^2}{2}\int d^4 x 
\partial_\mu\phi\partial^\mu\phi\,,
\label{goldstoneaction}
\end{eqnarray}
where the Goldstone coupling constant $f_\phi^2$ is defined as
\bq
f_\phi^2 = R^2 \int dy (\Phi_0')^2 = T R^2\,.
\label{fphi}
\eq 

 If one has a  multisoliton solution (for example, a system of
the  BPS
saturated    domain walls or D-branes) one has several Goldstone 
bosons
corresponding to independent positions   of these solitons.

Let us now take  gravity into account. Using the standard Kaluza-Klein
decomposition of the  metric (let us note that  we  use
the  angle $\phi$
as the fifth coordinate now)
\bq
ds^2 = g_{\mu\nu} dx^\mu dx^\nu + R^2(d\phi + A_\mu dx^\mu)^2
\label{metric}
\eq
one gets the  metric tensors $G_{MN}$ and $G^{MN}$, 
\begin{eqnarray}
 G_{MN}= \left( \begin{array}{cc}
g_{\mu\nu} + R^2 A_\mu A_\nu &  R^2 A_\mu\\
R^2 A_\nu & R^2\\ \end{array}\right), \qquad
 G^{MN}= \left( \begin{array}{cc} g^{\mu\nu} & -A^{\mu}\\
-A^{\nu} & R^{-2} + A_\mu A^\mu \\ \end{array}\right).
\label{KKmetric}
\end{eqnarray}
 
 It  easy to see that now  the global U(1) becomes local --
it  becomes  a  special  diffeomorphism
\bq
\phi  \ra  \phi  + \epsilon (x),\qquad x_\mu \ra x_\mu\,,
\label{localu1}
\eq
so that only the component $G_{\mu 4} = R^2 A_\mu$ 
is changed under this transformation. One can see 
that this is the case   from the expression for the 
general diffeomorphism
\bq
x^M \ra x^M + \epsilon^M,~~~~
G_{MN} \ra G_{MN} + \partial_M \epsilon_N + \partial_N\epsilon_M\,,
\eq
or, even easier, from (\ref{metric}) which immediately gives the gauge 
transformation of the
 vector field $A_\mu \ra A_\mu -
\partial_\mu\epsilon$.   In this paper we are not interested in the
$g_{\mu\nu}$ part of the metric. Even though in 
the  presence of a brane it
may be nontrivial and $y$ dependent it will not  affect the
properties of the classical solution for
the  scalar field itself (because  we
assume  that the scalar field depends only on
the  fifth coordinate and, therefore,   it
does not  matter how the four-dimensional metric  depends on $y$).

In the absence of  the brane one can get the
four-dimensional action by
substituting (\ref{KKmetric}) into  (\ref{5action}). Assuming that
the fields
do not depend on $y$  one gets the  four-dimensional 
Einstein-Hilbert plus Maxwell action  (we can call it the Kaluza-Klein
action) 
\bq
S_{KK} =  M_p^2\int d^4 x  \sqrt{g} R^{(4)} -\frac{1}{4 e^2} 
\int d^4 x \sqrt{g} F_{\mu\nu} F^{\mu\nu}
\label{KKaction}
\eq
where the Planck mass $M_p^2 = 2\pi R  M_{\rm P_{f}}^3$ and the
 U(1) coupling constant is
 $1/e^2 = M_p^2 R^2 = 2\pi ( M_{\rm P_{f}}R) ^3$.
 
 If we take into account the scalar action, there will be three terms, 
 \bq
 \frac{1}{2}G^{44}\partial_{y}\Phi_{0} \partial_{y}\Phi_{0}
 +G^{4\mu}\partial_{y}\Phi_{0} \partial_{\mu}\Phi_{0} +
 \frac{1}{2}G^{\mu\nu}\partial_{\mu}\Phi_{0} \partial_{\nu}\Phi_{0}\,.
 \eq
After taking into account that
 \bq
 \partial_{\mu}\Phi_{0} = - \partial_{y}\Phi_{0} R\partial_{\mu}\phi
 \eq
 and integrating over 
the fifth direction taking into account (\ref{tension})
  one gets a generalization of (\ref{goldstoneaction}), so that instead
of 
  $\partial_\mu\phi$ we have $\partial_\mu\phi + A_\mu$ 
and the combined   action
\bq
S =   -\frac{1}{4 e^2} \int d^4 x \sqrt{g} F_{\mu\nu} F^{\mu\nu} +
\frac{1}{2}f_\phi^2 \int d^4 x \sqrt{g}
(\partial_\mu\phi + A_\mu)(\partial^\mu\phi + A^{\mu})\,.
\label{higgsaction}
\eq 
The latter tells us that graviphoton becomes a massive vector 
particle with the mass
\bq
  M^2_V = f_\phi^2 e^2 = \frac{T}{M_p^2}\,.
 \eq

\end{document}